\documentstyle[sprocl]{article}

\input{psfig}
\def\be{\begin{equation}}
\def\ee{\end{equation}}
\def\bea{\begin{eqnarray}}
\def\eea{\end{eqnarray}}

\newcommand{\pom}{I\!\! P}
\begin{document}

\title{DIFFRACTION IN CDF:\\
RUN I RESULTS AND PLANS FOR RUN II 
\footnote{Presented at DIS-2001,
Bologna, Italy, 27 April - 1 May 2001.}
}
\author{K. GOULIANOS}

\address{(For the CDF Collaboration)\\
The Rockefeller University, 1230 York Avenue, New York,\\ 
NY 10021, USA\\E-mail: dino@physics.rockefeller.edu}




\maketitle\abstracts{
	Results on diffraction obtained by the CDF Collaboration 
	in Run I of the Fermilab Tevatron $\bar pp$ collider are reviewed.
	New results are reported on soft double diffraction and 
	diffractive $J/\psi$ production.
	The CDF program for diffractive studies in Run II is briefly 
	discussed.
}  
\vspace*{-9.5cm}
\phantom{xxx}
\vspace*{8.25cm}
\section{Introduction}
\vspace*{-0.25cm}
The signature of a diffractive event in $\bar pp$ collisions is a
leading proton or antiproton and/or a rapidity gap, defined as a region of 
pseudorapidity, $\eta\equiv -\ln\tan\frac{\theta}{2}$, devoid of particles 
(see Fig.~1). 

\begin{figure}[h,t]
\vglue -0.5cm
\centerline{\hspace*{1cm}\psfig{figure=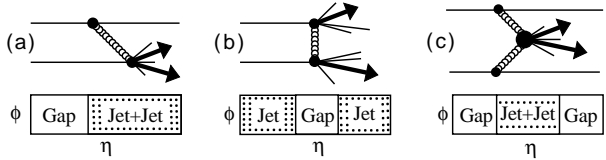,width=4.2in}}
\vglue -4.7in
\caption{Dijet production diagrams and event topologies for 
(a) single-diffraction,  (b) double-diffraction, and
(c)  double Pomeron exchange.}
\label{topology}
\end{figure}

In Run I, CDF studied the following diffractive processes:
\begin{itemize}
\item soft single-diffraction (SD) at $\sqrt s=$ 546 and 1800 GeV~\cite{CDF_SD} 
and soft double-diffraction (DD) at $\sqrt s=$ 630 
and 1800 GeV~\cite{CDF_DD}
\item $W$-boson~\cite{CDF_W}, dijet~\cite{CDF_JJG}, 
$b$-quark~\cite{CDF_b} and $J/\psi$~\cite{CDF_jpsi} production 
at $\sqrt s=$ 1800 GeV using rapidity gaps to identify diffractive events
\item dijets with a rapidity gap between jets at $\sqrt s=$ 
630~\cite{CDF_JGJ630} and 1800 GeV~\cite{CDF_JGJ1800_1,CDF_JGJ1800_2}
\item dijets with a leading antiproton 
at $\sqrt s=$ 630~\cite{CDF_JJ630} and 1800 GeV~\cite{CDF_JJ1800}
\item double Pomeron exchange (DPE) dijet production with a leading 
antiproton and a rapidity gap on the proton side~\cite{CDF_DPE}
\end{itemize}
In our discussion below,  we address the issues of 
universality in rapidity gap formation and of Regge and QCD factorization
in hard diffraction.
\newpage
\section{Soft diffraction}
Measurements of $pp$ and $\bar pp$ SD cross sections have shown that Regge 
theory correctly predicts the shape of the rapidity gap dependence for 
$\Delta\eta>3$, corresponding to a leading proton fractional momentum loss of 
$\xi=e^{-\Delta\eta}<0.05$, but fails to predict the 
correct energy dependence of the overall normalization, which 
at $\sqrt s=1800$ GeV is found to be suppressed by approximately an order of
magnitude~\cite{CDF_SD,dino_flux,GM}. A new CDF measurement of the double 
diffraction differential cross section gives similar results (see Fig.~2).
\vglue -1em
\begin{minipage}[t]{2.1in}
\phantom{x}
\vglue -1cm
\hspace*{-1cm}\psfig{figure=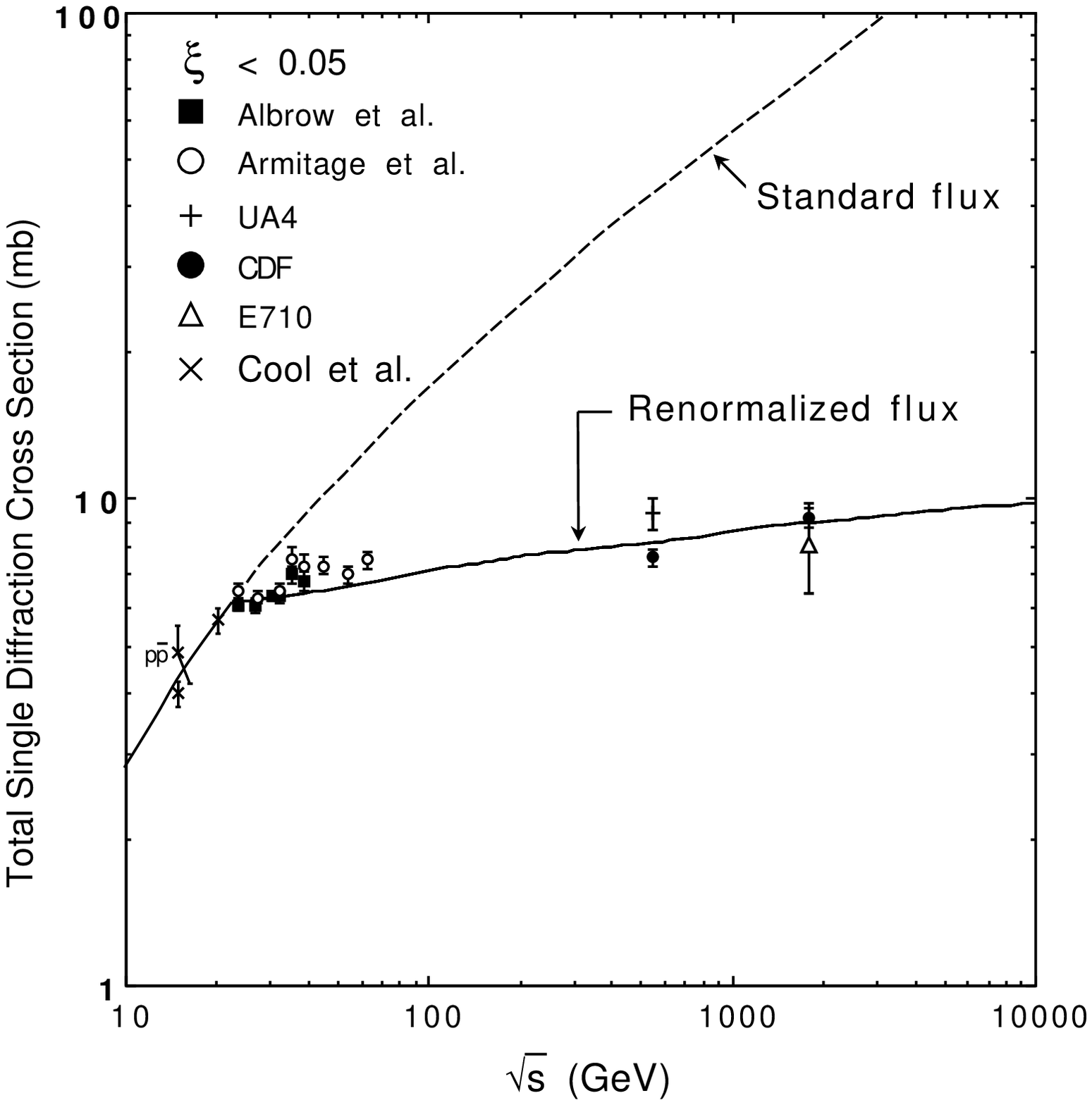,width=2.60in}
\vglue -2cm
{\small Figure 2a: The $pp/\bar pp$ $\sigma^T_{SD}$ versus $\sqrt s$.}
\end{minipage}
\hspace*{0.1cm}
\begin{minipage}[t]{2.25in}
\phantom{x}
\vglue -0.25cm
\psfig{figure=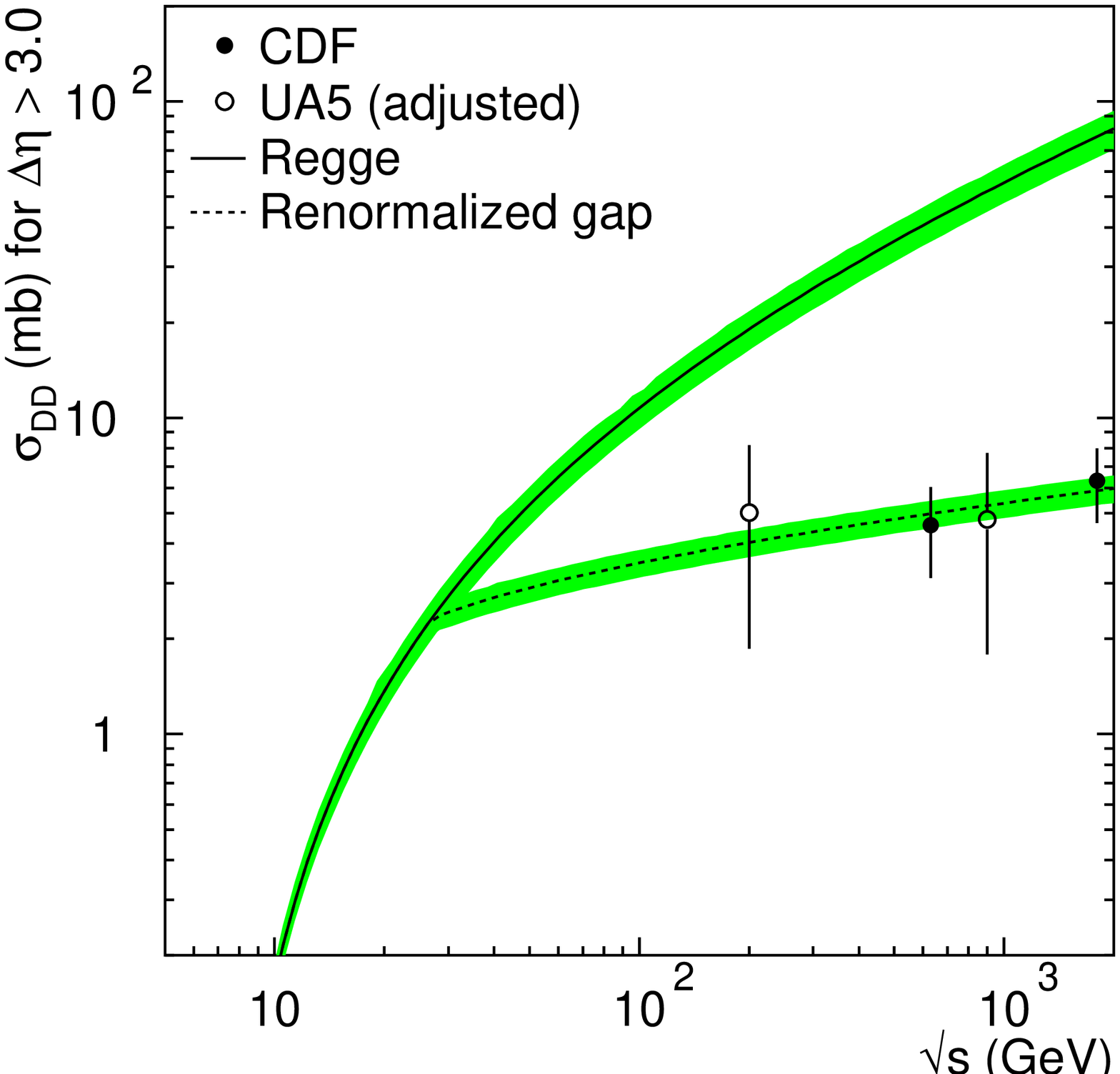,width=2.35in}
{\hspace*{0.8cm}{\small Figure 2b: The $\bar pp$ $\sigma^T_{DD}$ 
versus $\sqrt s$.}}
\vglue -1.9cm
\hspace*{3cm}{\bf\large\it Preliminary}
\end{minipage}

\vglue 0.2cm

The SD and DD cross sections have very similar forms in terms of $\Delta\eta$:
\vspace*{-0.2cm}
\begin{eqnarray}
d^2\sigma_{SD}/dt\,d\Delta\eta=&
[Ke^{bt}e^{[2\alpha(t)-1]\Delta\eta}]\cdot 
[\kappa \beta^2(0)(s')^{\alpha(0)-1}]\\
d^3\sigma_{DD}/dt\,d\Delta\eta\,d\eta_c=&
[\kappa \;K\,e^{[2\alpha(t)-1]\Delta\eta}]
\cdot[\kappa \beta^2(0)(s')^{\alpha(0)-1}]
\end{eqnarray}
\vglue -0.2cm
Here, energy is measured in GeV,
$\alpha(t)=\alpha(0)+\alpha't$ is the Pomeron trajectory, 
$\beta(t)$ is the coupling of the Pomeron to the proton, $K=\beta^2(0)/16\pi$,
$\kappa=g_{\pom\pom\pom}/\beta(0)$, where $g_{\pom\pom\pom}$
is the triple-Pomeron coupling, $e^{bt}$ is the square of the proton 
form factor, $\eta_c$ the center of the rapidity gap, and  
$s'=M_1^2M_2^2$ ($M$ is the diffractive mass) can be thought of as the 
$s$-value of the diffractive sub-system(s), since $\ln s'=\ln s-\Delta\eta$ 
is the rapidity space where particle production occurs.
The second factor in the equations 
can be thought of as the sub-energy total cross section, which allows the 
first factor to be interpreted as a rapidity gap probability, $P_{gap}$.
For SD, it has been shown that renormalizing the Pomeron flux~\cite{dino_flux}, 
which is equivalent to normalizing $P_{gap}$ over all phase space 
to unity, yields the correct energy dependence. The new CDF results
show that this also holds for DD, as predicted 
by a generalization of the Pomeron flux renormalization model~\cite{dino_gap}.
\newpage
\section{Hard diffraction using rapidity gaps}
Using forward rapidy gaps to tag diffractive events, CDF 
measured the ratio of SD to non-diffractve (ND) rates for 
$W$-boson~\cite{CDF_W}, dijet~\cite{CDF_JJG}, 
$b$-quark~\cite{CDF_b} and $J/\psi$~\cite{CDF_jpsi} production 
at $\sqrt s=$ 1800 GeV, and using central gaps determined the fraction of 
jet-gap-jet events as a function of $E_T^{jet}$ and of rapidity gap separation 
between the two jets ($\Delta\eta^{jet}$) at $\sqrt s=$ 630 and 1800 GeV.

Forward gaps were defined as no hits in one of the beam-beam counters, BBC
($3.2<|\eta|<5.9$), and no towers with energy
$E>1.5$ GeV in the forward calorimeters, FCAL ($2.4<|\eta|<4.2$). Using the 
POMPYT Monte Carlo (MC) simulation with a flat gluon/quark Pomeron structure,
the measured SD/ND ratios were corrected for `gap acceptance', defined as 
the ratio of diffractive events with a gap to all diffractive events 
generated with $\xi=x_{\pom}<0.1$ in the selected kinematical range 
of the hard scattering products. 

For jet-gap-jet events, the gap was defined as no tracks or towers with 
energy above $\sim 300$ MeV in the region $|\eta|<1$.
The ND background was estimated using events with both jets 
at positive or negative $\eta$.
\vglue 1em
\centerline{\small Table 1: Ratios of diffractive ($\xi<0.1$) to 
non-diffractive rates.}
\vglue -1in
\centerline{\psfig{figure=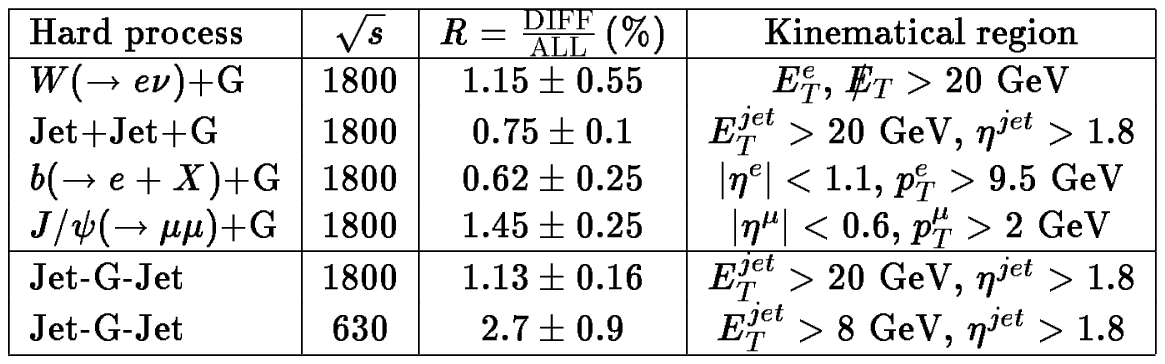}}
\vglue -8.45in

The results are summarized in Table~1. At $\sqrt s $=1800 GeV 
the DIFF/ALL ratios  
are approximately equal. Since the processes under study have different 
sensitivities to the quark and gluon content of the Pomeron, these results 
indicate that the value of the gluon fraction in the Pomeron, $f_g^{\pom}$, 
is not very different from that in the proton. From the $W$, dijet and 
$b$-quark ratios, $f_g^{\pom}$ was determined to be~\cite{CDF_b} 
$0.54^{+0.16}_{-0.14}$. In addition, a suppression of a factor 
$D=0.19\pm 0.04$ was found in these ratios relative to POMPYT 
predictions using the standard Pomeron flux. This discrepancy indicates a 
breakdown of factorization~\cite{CDF_b}. The value of 
$D$ is approximately the same as that in soft SD (see Fig.~2), 
as predicted in Ref.~\cite{dino_flux}.

The ratio of jet-gap-jet fractions at $\sqrt s=630$ to 1800 GeV is 
$2.4\pm 0.8$. The $\Delta\eta^{jet}$, $E_T^{jet}$ and $x$-Bjorken
distributions are consistent with being flat~\cite{CDF_JGJ1800_2}. 
\newpage
\section{Hard diffraction using a leading antiproton spectrometer}
Using a Roman pot spectrometer to detect leading antiprotons 
and determine their momentum and polar angle (hence the $t$-value),
CDF measured the ratio of SD to ND dijet production rates 
at $\sqrt s$=630~\cite{CDF_JJ630} and 1800 GeV~\cite{CDF_JJ1800} as a 
function of $x$-Bjorken of the struck parton in the $\bar p$. In leading order
QCD, this ratio is equal to the ratio of the corresponding 
structure functions. For dijet production, the relevant structure function is
the color-weighted combination of gluon and quark terms given by 
$F_{jj}(x)=x[g(x)+\frac49\sum_iq_i(x)]$. The diffractive structure function,
$\tilde{F}_{jj}^D(\beta)$, where $\beta=x/\xi$ is the momentum fraction of the 
Pomeron's struck parton, is obtained by multiplying 
the ratio of rates by the known $F_{jj}^{ND}$ and changing variables 
from $x$ to $\beta$ using $x\rightarrow \beta\xi$  
(the tilde over the $F$ indicates integration over 
$t$ and $\xi$, as specified in each case). 

Results for $\sqrt s=$ 1800 GeV are presented in Fig.~3: 

\begin{minipage}[t]{2.5in}
\phantom{x}
\vglue -0.5cm
\hspace*{-1cm}\psfig{figure=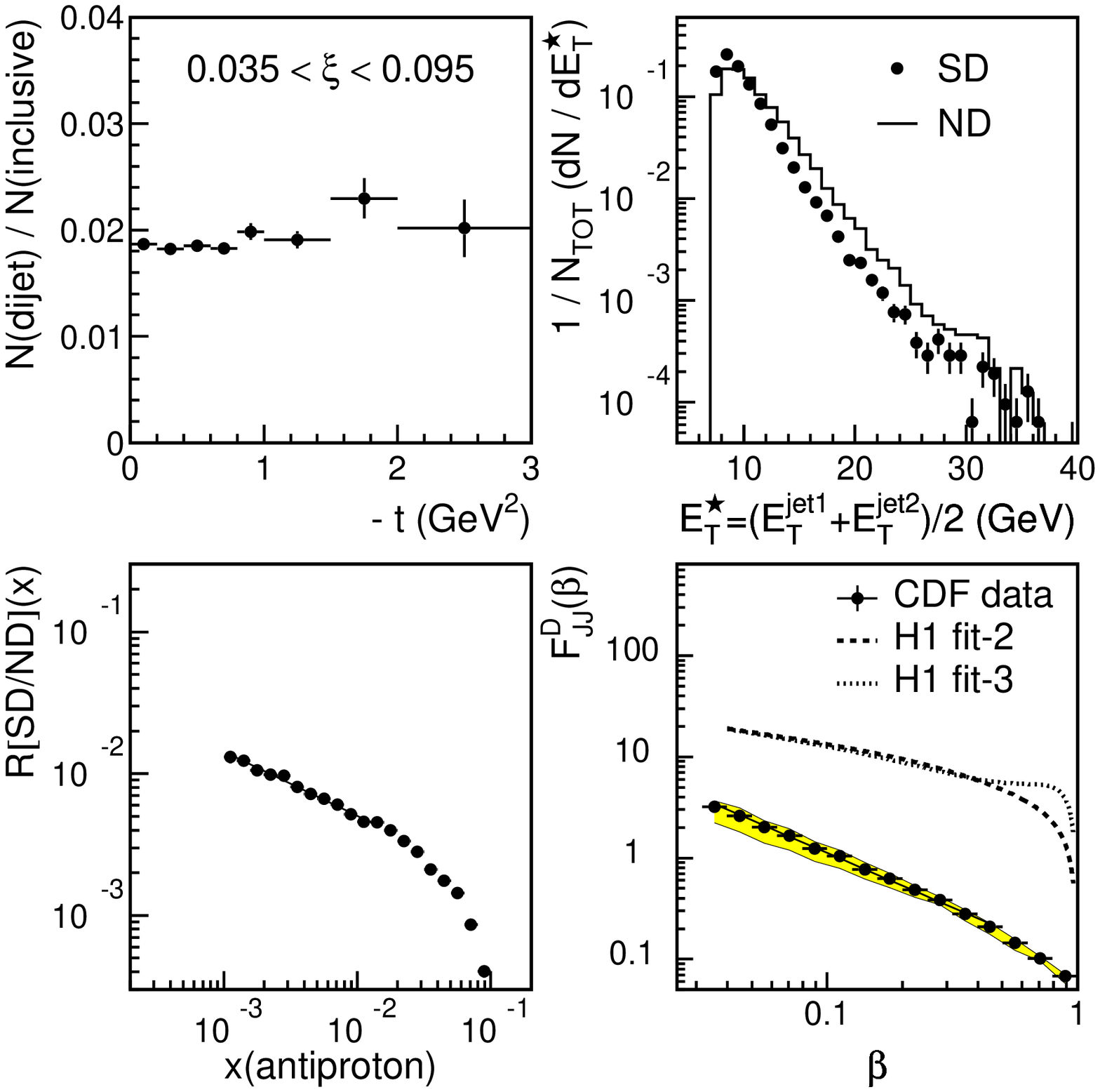,width=2.5in}
\end{minipage}
\hspace*{-1.2cm}
\begin{minipage}[t]{2.4in}
\phantom{x}
{\small Figure 3: Results from inclusive and dijet 
diffractive data with a leading $\bar p$:\\
{\em (top left)} the ratio 
of dijet to inclusive SD event rates is independent of $t$;\\
{\em (top right)}
the $E_T^{jet}$ distribution is slightly steeper for SD than for ND events;\\
{\em (bottom left)} the ratio of SD to ND rates increases with decreasing 
$x_{bj}$;\\
{\em (bottom right)} the CDF diffractive structure function 
is steeper than and severely suppressed relative to 
predictions based on extrapolations of the diffractive parton densities 
extracted by the H1 Collaboration from DIS measurements at HERA.}
\end{minipage}
\vglue 1em
The above results confirm the breakdown of factorization observed in the 
rapidity gap data presented in section 3. Differences in suppression 
factors can be traced back to differences in kinematical acceptance.

Factorization was also tested {\em within CDF data} by comparing the 
ratio of DPE to SD to that of SD to ND dijet production rates.
The DPE events were extracted from the leading antiproton 
data by demanding a rapidity gap in the forward detectors on the 
proton side, defined as in section 3. At $\langle \xi\rangle=0.02$ and 
$\langle x_{bj}\rangle =0.005$, the ratio of SD/ND to DPE/SD rates 
normalized per unit $\xi$ was found to be~\cite{CDF_DPE} $0.19\pm 0.07$,
violating factorization. 

	A search for the process 
$p+\bar p\rightarrow p'+(jet1+jet2)+\bar p'$ yielded~\cite{CDF_DPE} 
an upper limit 
of 3.7 nb for $0.035<\xi(\bar p)<0.095$ and jets of $E_T>7$ GeV and 
$\eta<1.7$.
\newpage
\section{Plans for Run II}
The CDF program for diffractive studies in Run II will include:\\
{\large (a) Hard single diffraction}\\
\hspace{1cm}-- Process dependence of $F^D$ (compare at the same $\xi$ and 
$x_{bj}$)\\ 
\hspace{1cm}-- $Q^2$ dependence of $F_{jj}^D$\\
{\large (b) Double Pomeron exchange}\\
\hspace{1cm}-- Soft DPE\\
\hspace{1cm}-- $F_{jj}^D(x_p)$ versus width of gap on the $\bar p$ side\\
\hspace{1cm}-- Exclusive dijet and $\bar bb$ production\\
\hspace{1cm}-- Low mass exclusive states (glueballs?)\\
{\large (c) Hard double diffraction}\\
\hspace{1cm}-- jet-gap-jet events at high $\Delta\eta^{jet}$ (test BFKL)\\
{\large (d) Unexpected discoveries!}
\vglue 0.1in
The Run II program will be implemented by upgrading CDF to include the forward 
detector system shown schematically in Fig.~4. This system comprises:
1. A Roman Pot Specrometer (RPS) on the antiproton side
to detect leading antiprotons and measure $\xi$ and $t$\\
2. {Beam Shower Counters (BSC) covering the region $5.5<|\eta|<7.5$
to be used for triggering on events with forward rapidity gaps}\\
3. {Two `MiniPlug' calorimeters in the region $3.5<|\eta<5.5$}\\
\vglue -0.35in
\centerline{\psfig{figure=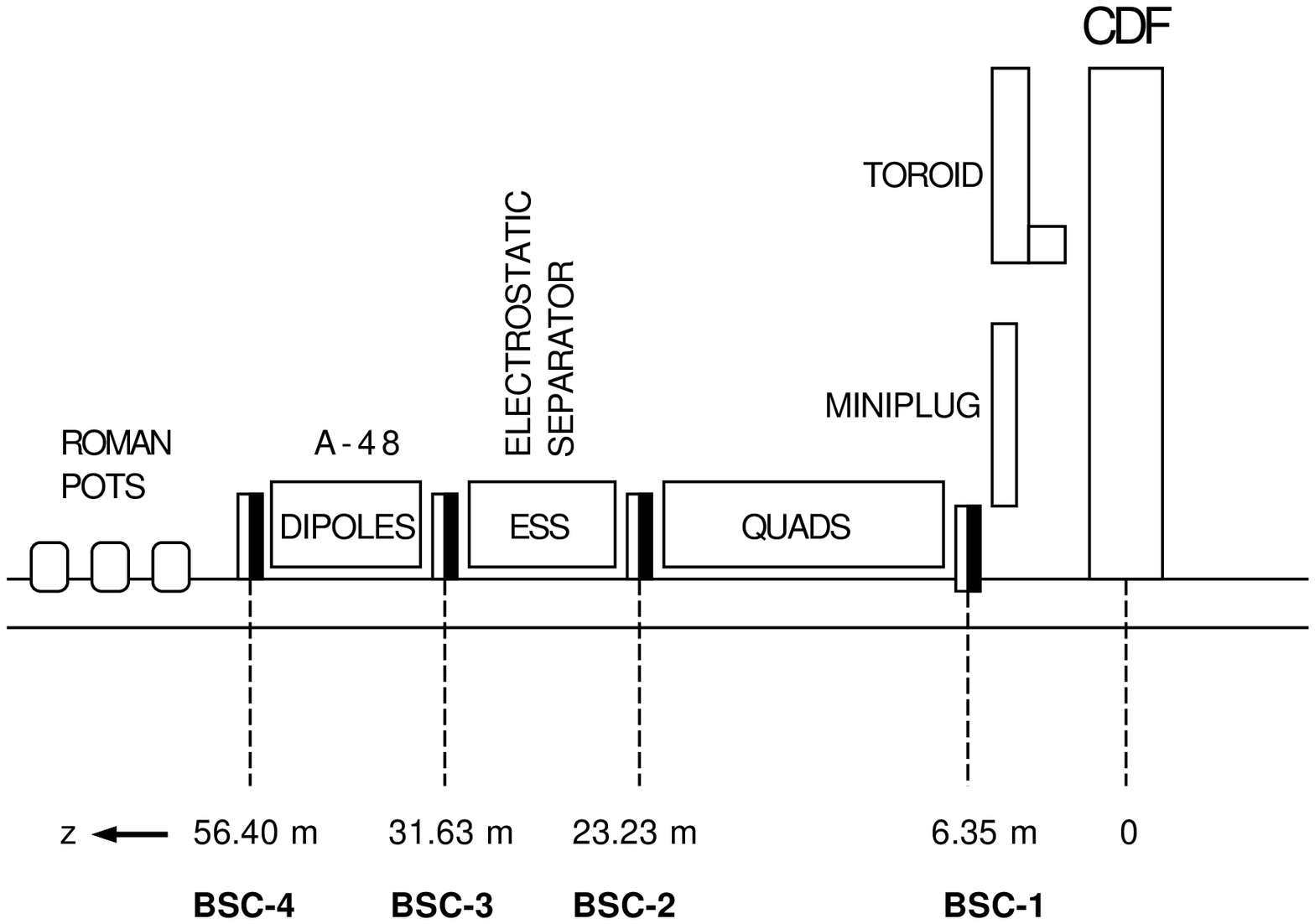,width=3.5in}}
\vglue -2in
\centerline{\small Figure 4: CDF forward detectors for Run II}
\vglue 0.1cm
The RPS and BSC systems are already installed, and the MiniPlug 
installation is scheduled for September 2001. 


\newpage
\section*{References}
\begin{thebibliography}{99}
%
\bibitem{CDF_SD}F. Abe {\em et al.}, Phys. Rev. D {\bf 50}, 5535 (1994).
%
\bibitem{CDF_DD}{\em Double diffraction dissociation at the Fermilab Tevatron 
collider}, CDF Collaboration, submitted to Phys. Rev. Letters.
%
\bibitem{CDF_W}F. Abe {\em et al.}, Phys. Rev. Lett. {\bf 78}, 2698 (1997).
%
\bibitem{CDF_JJG}F. Abe {\em et al.}, Phys. Rev. Lett. {\bf 79}, 2636 (1997).
%
\bibitem{CDF_b}T.~Affolder {\em et al.}, Phys. Rev. Lett. {\bf 84}, 232 (2000).
%
\bibitem{CDF_jpsi}{\em Observation of diffractive $J/\psi$ production at 
the Fermilab Tevatron}, CDF Collaboration, to be submitted to
Phys. Rev. Letters.
%
\bibitem{CDF_JGJ630} F. Abe {\em et al.},
Phys. Rev. Lett. {\bf 80}, 1156 (1998); {\bf 81}, 5278 (1998).
%
\bibitem{CDF_JGJ1800_1}F. Abe {\em et al.}, 
Phys. Rev. Lett. {\bf 74}, 855 (1995).
%
\bibitem{CDF_JGJ1800_2} F. Abe {\em et al.}, 
Phys. Rev. Lett. {\bf 80}, 1156 (1998).
%
\bibitem{CDF_JJ630}{\em Diffractive dijet production at $\sqrt s=630$ and 1800 
GeV at the Fermilab Tevatron}, CDF Collaboration, to be submitted to 
Phys. Rev. Letters.
%
\bibitem{CDF_JJ1800}T.~Affolder {\em et al.},
Phys. Rev. Lett. {\bf 84}, 5043 (2000).
%
\bibitem{CDF_DPE} T.~Affolder {\em et al.},
Phys. Rev. Lett. {\bf 85}, 4215 (2000).
%
\bibitem{dino_flux}K. Goulianos, Phys. Lett. B 358, 379 (1995).
%
\bibitem{GM}K. Goulianos and J. Montanha,
Phys. Rev. D {\bf 59}, 114017 (1999).
%
\bibitem{dino_gap}K. Goulianos, ``Diffraction: Results and Conclusions",
in {\em Proceedings of
Lafex International School of High Energy Physics, Rio de Janeiro, Brazil,
February 16-20 1998}, edited by
Andrew Brandt, H\'{e}lio da Motta and Alberto Santoro; hep-ph/9806384.
\end{thebibliography}
\end{document}